\documentclass[10pt,reqno]{article}
\usepackage{fullpage}

\usepackage[unicode=true]{hyperref}
\usepackage{amsmath}
\usepackage{cite}
\usepackage{amsfonts}
\usepackage{amssymb}
\usepackage{amsthm}
\usepackage{authblk}

\usepackage[pdftex]{color,graphicx}
\usepackage{mypack} 
\usepackage{adjustbox}
\usepackage{soul}
\usepackage{comment}
\usepackage{bbold}
\usepackage{tikz}
\usetikzlibrary{decorations.pathreplacing,angles,quotes}
\usepackage{bbold}

\usepackage{diagbox}
\usepackage{enumerate}

\usetikzlibrary{matrix,arrows,decorations.pathmorphing}
\usepackage{tikz-cd}
\usetikzlibrary{arrows} 

\usetikzlibrary{decorations.markings}
 \usepackage{caption}
\usepackage{subcaption}
 
 \usepackage{pdfpages}

\usepackage{blkarray}

\usepackage{centernot}
\usepackage{mathtools}
\usepackage{stmaryrd}

  \usepackage{arydshln}

\definecolor{bluegray}{rgb}{0.4, 0.6, 0.8}
\definecolor{turquoise}{rgb}{0.2, 0.7, 0.6}
\definecolor{hy-green}{rgb}{0.1, 0.5, 0.1}

\usepackage{soul}  
 \newcommand{\suchthat}{\;\ifnum\currentgrouptype=16 \middle\fi|\;}

\title{On the rank of two-dimensional simplicial distributions}

\author{Cihan Okay\footnote{cihan.okay@bilkent.edu.tr}} 
\affil{Department of Mathematics, Bilkent University, Ankara, Turkey} 

\begin{document}
  \maketitle

\begin{abstract}
Simplicial distributions provide a framework for studying quantum contextuality, a generalization of Bell's non-locality.
Understanding extremal simplicial distributions is of fundamental importance with applications to quantum computing. 
We introduce a rank formula for twisted simplicial distributions defined for $2$-dimensional measurement spaces and provide a systematic approach for describing extremal distributions.
\end{abstract}

\tableofcontents

\section{Introduction} 

{The theory of simplicial distributions}
introduced in \cite{okay2022simplicial} is a framework for describing quantum contextuality, a fundamental feature of quantum theory generalizing Bell's non-locality \cite{bell64,KS67}.
In this framework, measurements and outcomes are represented by spaces modeled by combinatorial objects called simplicial sets \cite{goerss2009simplicial}.
This framework generalizes the theory of non-signaling distributions formulated in the language of sheaf-theory \cite{abramsky2011sheaf} 
by elevating {\it sets} of measurements and outcomes to {\it spaces} of measurements and outcomes.
Simplicial distributions constitute a polytope whose vertices (extreme distributions)  are of fundamental importance in quantum foundations \cite{pitowsky1989quantum,popescu1994quantum,barrett2005nonlocal,
jones2005interconversion}. Non-contextual distributions are described by a subpolytope whose facets are given by the Bell inequalities. In this paper, we introduce graph-theoretic methods to identify the contextual vertices of the polytope of twisted simplicial distributions.

\begin{figure}[h!]
\centering
\begin{subfigure}{.49\textwidth}
  \centering
  \includegraphics[width=.4\linewidth]{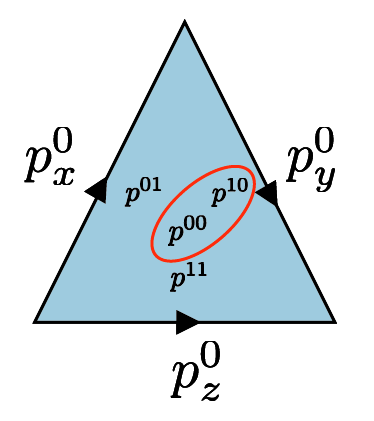}
  \caption{}
  \label{fig:two-simplex-distribution}
\end{subfigure}%
\begin{subfigure}{.49\textwidth}
  \centering
  \includegraphics[width=.4\linewidth]{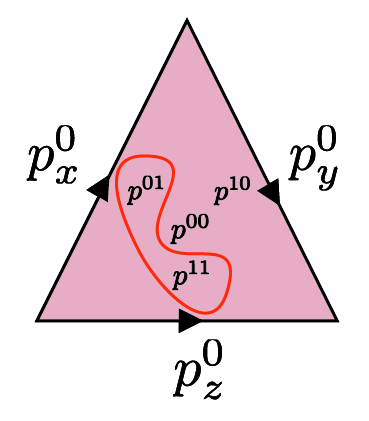}
  \caption{}
  \label{fig:two-simplex-twisted-distribution}
\end{subfigure}
\caption{Triangle $\sigma$ with faces $x=d_2\sigma$, $y=d_0\sigma$ and $z=d_1\sigma$. (a) A twisted simplicial distribution on $\sigma$ with $\beta(\sigma)=0$. The $d_0$-face is given by $p_y^0=p^{00}+p^{10}$. (b) A twisted simplicial distribution  with $\beta(\sigma)=1$. The $d_0$-face is given by $p_y^0=p^{01}+p^{11}$.}
\label{fig}
\end{figure}

A simplicial distribution on a scenario consisting of a space $X$ of measurements and space $Y$ of outcomes is a simplicial set map of the form 
$$
p:X\to D(Y)
$$ 
where $D(Y)$ is a simplicial set that models the space of distributions on the outcome space.
In this paper we will restrict to the case where $X$ is a simplicial set generated by   $2$-dimensional simplices $\sigma_1,\sigma_2,\cdots,\sigma_N$, and $Y$ is the nerve space of the additive group $\ZZ_2=\set{0,1}$. Topologically $X$ is a $2$-dimensional space obtained by gluing the $2$-simplices (triangles) along their faces and possibly collapsing some of them.
Our choice of the outcomes space reflects the restriction that our measurements have binary outcomes. More concretely, a simplicial distribution $p$ consists of a family $\set{p_{\sigma_i}:\,i=1,2,\cdots,N}$ of probability distributions 
$$
\set{p_{\sigma_i}^{ab}\in \RR_{\geq 0}:\,a,b\in \ZZ_2,\, \sum_{a,b} p_\sigma^{ab}=1}$$ 
satisfying a compatibility condition induced by the face relations. Simplicial distributions on $X$ define a polytope with finitely many vertices.
The face relations impose conditions in the form of marginals, e.g., for the $d_0$-face of a triangle $\sigma$, we have
$$
p_{d_0\sigma}^0 = p^{00}_\sigma+p^{10}_\sigma. 
$$
In this paper, we extend our interest to twisted distributions. For a $2$-cocycle $\beta:X_2\to \ZZ_2$ we write $\sDist_\beta(X)$ for the polytope of $\beta$-twisted simplicial distributions on $X$.
The effect of the twisting appears only in the $d_0$-face, and the marginalization formula above is generalized as follows:
$$
p_{d_0\sigma}^0 = 
\left\lbrace
\begin{array}{ll}
p^{00}_\sigma+p^{10}_\sigma & \beta(\sigma) =0\\
p^{11}_\sigma+p^{01}_\sigma & \beta(\sigma) =1.
\end{array}
\right.
$$
Given a twisted simplicial distribution $p$, we define a simplicial subset $Z_p\subset X$ consisting of simplices on which the distribution is deterministic, i.e., given by a delta-distribution. 
The rank of $p$ is defined to be the rank of the matrix consisting of the defining inequalities of the polytope $\sDist_\beta(X)$ that are tight at $p$.
Our main result provides a formula for the rank. {For a simplicial set $X$, w}e will write {$X_n^\circ$} to denote the subset of non-degenerate simplices.
Given a triangle $\sigma$, the set of 
{$1$-simplices} in its boundary is denoted by $\partial \sigma$.

\begin{thm*}  
Let $X$ be a simplicial set 
generated by   $2$-simplices $\sigma_1,\sigma_2,\cdots ,\sigma_N$ {such that
each $\partial\sigma_i$ consists of either three distinct non-degenerate {$1$-simplices} or two distinct non-degenerate {$1$-simplices} and a remaining degenerate {$1$-simplex}.}
{Consider}  
a twisted distribution $p\in \sDist_\beta(X)$ satisfying the following conditions:
\begin{itemize}
\item  for each generating $2$-simplex $\sigma$, $p_\sigma^{ab}=0$ for at least one pair $(a,b)\in \ZZ_2^2$, and
\item every non-degenerate $1$-simplex of $\bar X=X/Z_p$ belongs  precisely to two generating $2$-simplices.
\end{itemize}
Then we have
$$
\rank(p) = |(Z_p)_1^\circ| + |\bar X_2^\circ| - b(\bar X,\bar p).
$$ 
\end{thm*}

The crucial component of the rank formula is $b(X,p)$, a natural number defined for a twisted simplicial distribution. Our graph-theoretic approach is based on constructing a signed graph associated with $p$ and $b(X,p)$ is the number of balanced components in this graph.
Balancedness is an important property for signed graphs.
A connected graph is called balanced if the sign of every circle contained in it, which is simply given by the product of the signs of the edges in the circle, is positive. In the theorem, we apply this construction to the quotient space $\bar X = X/Z_p$ and the twisted distribution $\bar p$ constructed in Proposition \ref{pro:zeta-twisted} using a cocycle obtained by a cohomology long exact sequence. In the last section, we demonstrate how to apply the rank formula to describe the vertices of $\sDist(X)$ for the following important scenarios:
\begin{itemize}
\item In Section \ref{sec:N-cycle-scenario}, we study cycle scenarios whose measurement space consists of a disk triangulated into $N$ triangles. This scenario is the generalization of the famous Clauser--Horne--Shimony--Holt  (CHSH) scenario \cite{clauser1969proposed}.

\item The scenario whose measurement space is the boundary of a tetrahedron is considered in Section \ref{sec:boundary of tetrahedron}. Describing these vertices was a key step in the topological proof of Fine's theorem \cite{okay2022simplicial}, that characterizes non-contextual distributions on the CHSH scenario.

\item The Mermin scenario \cite{mermin1993hidden} provides contextual distributions that arise in quantum theory. In \cite{Coho}, a topological realization is provided where the underlying measurement space is a torus. The vertices of the polytope of twisted distributions for different cocycles are studied in \cite{okay2022mermin}. We reproduce the vertices in Section \ref{sec:mermin torus}.
\end{itemize}
As displayed in these examples, the rank formula provides a systematic study of the vertices of the polytope of twisted distributions. We expect this approach will also be useful in analyzing more complicated polytopes in the context of classical simulation algorithms for quantum computation \cite{zurel2020hidden}. Some ideas in this direction appear in \cite{okay2022mermin,classint}.

The paper is structured as follows. In Section \ref{sec:simplicial distributions}, we introduce simplicial sets focusing on the $2$-dimensional case. We introduce twisted simplicial distributions in this restricted setting. Our main result of this section, Proposition \ref{pro:zeta-twisted}, is proved using ideas from  cohomology of simplicial sets and products of twisted distributions. In Section \ref{sec:Distributions on graphs}, we develop our graph-theoretical methods and introduce distributions on graphs of interest. Our rank formula is proved in Theorem \ref{thm:main}.
Section \ref{sec:Examples} contains the examples of practical interest where we put the rank formula in work.

\paragraph{Acknowledgments.}
This work is supported by the US Air Force Office of Scientific Research under award number FA9550-21-1-0002 and the Digital Horizon Europe project FoQaCiA, GA no. 101070558. 
{The author would like to thank Selman Ipek and Walker H. Stern for their helpful comments on an earlier version of this paper.}

\section{Simplicial distributions}
\label{sec:simplicial distributions}

In this section we recall some basic definitions from \cite{okay2022simplicial} on simplicial distributions and introduce a twisted version in the case of $2$-dimensional simplicial sets based on the exposition in  \cite{barbosa2023bundle}. 
Our main result is Proposition \ref{pro:zeta-twisted}, an important ingredient for the rank formula in Section \ref{sec:Distributions on graphs}.

\subsection{Two-dimensional simplicial sets}
\label{sec:Two-dimensional simplicial sets}

A simplicial set $X$ consists of a sequence of sets $X_n$ for $n\geq 0$ together with the face maps $d_i:X_n\to X_{n-1}$ and the degeneracy maps $s_j:X_n\to X_{n+1}$ satisfying the simplicial identities \cite{goerss2009simplicial,friedman2008elementary}.
A simplicial set map $f:X\to Y$ consists of functions $f_n:X_n\to Y_n$ for $n\geq 0$ compatible with the face and the degeneracy maps. We will write $f_\sigma$ for the simplex $f_n(\sigma)\in Y_n$ for a given $n$-simplex $\sigma\in X_n$.
With this notation compatibility with the simplicial structure can be expressed as
$$
d_i f_\sigma = f_{d_i \sigma}\;\;\text{ and }\;\;
s_j f_\sigma = f_{s_j\sigma}.
$$
{A simplex is called non-degenerate if it does not lie in the image of a degeneracy map. A non-degenerate simplex is called generating if it does not lie in the image of a face map. For a set $U$ of simplices we write $U^\circ$ to denote the subset of non-degenerate simplices.}

{In this paper we only consider simplicial sets where each $X_n$ is a finite set. We will also have a restriction on the dimension of the simplicial set.}
A $1$-dimensional simplicial set is the same as a directed graph  with vertex set $X_0$ and edge set $X_1$ {(allowing loops and parallel edges)}. A $1$-simplex can be treated as an arrow
$$
v\xrightarrow{x} w
$$
with source $d_1(x)=v$ and target $d_0(x)=w$. The degenerate $1$-simplex $s_0(v)$ have both source and target given by $v$. 
A $2$-dimensional simplicial set consists of such a directed graph specified by $(X_0,X_1,d_0,d_1,s_0)$ together with a finite set of $2$-simplices $\sigma_1,\cdots,\sigma_N$  glued to the graph. The gluing is encoded by specifying the faces $d_i(\sigma_k) \in X_1$ for $i=0,1,2$.   
For example, the standard $2$-simplex $\Delta^2$ has a single generating $2$-simplex $\sigma$ glued to the directed graph on the three vertices $\set{0,1,2}$ with edges $0\to 1$, $1\to 2$ and $0\to 2$.
The face maps are given by
$$
d_i\sigma = \left\lbrace
\begin{array}{ll}
1\xrightarrow{y} 2 & i=0\\
0\xrightarrow{z} 2 & i=1\\
0\xrightarrow{x} 1 & i=2.
\end{array}
\right.
$$
The degenerate edges $0\xrightarrow{s_0(0)}0$, $1\xrightarrow{s_0(1)}1$ and $2\xrightarrow{s_0(2)}2$ are usually omitted. We will write $\partial \sigma=\set{x,y,z}$ for the set of edges in the boundary.
Let $f:X\to Y$ be a simplicial set map where $X$ is generated by the $2$-simplices $\sigma_1,\sigma_2,\cdots,\sigma_N$.
Such a map assigns a $2$-simplex $f_{\sigma_k}\in Y_2$ for each $k=1,\cdots,N$ such that
$$
d_i f_{\sigma_k} = d_j f_{\sigma_l}  
$$
whenever $d_i\sigma_k = d_j\sigma_l$. Therefore to study such maps it suffices to understand the face maps of $Y$ in dimension $2$:
$$
d_i:Y_2\to Y_1, \;\;\;\; i=0,1,2. 
$$
For simplicity when we introduce a simplicial set which will appear only in the target of a simplicial set we will only specify these face maps.

As an important target space we will be considering the nerve of the additive group $\ZZ_2=\set{0,1}$ of integers modulo $2$.
This is a simplicial set denoted by $N\ZZ_2$ whose $n$-simplices are given by $n$-tuples of elements in $\ZZ_2$. The face maps in dimension $2$ are given by
\begin{equation}\label{eq:faces of nerve}
d_i(a,b) = \left\lbrace
\begin{array}{ll}
b & i=0 \\
a+b & i=1 \\
a & i=2. 
\end{array}
\right.
\end{equation}

{
\Pro{\label{pro:map to nerve}
Let $X$ be a simplicial set with generating $2$-simplices $\sigma_1,\sigma_2,\cdots,\sigma_N$. There is a bijection between simplicial set maps $s:X\to N\ZZ_2$ and functions $\varphi:X_1\to \ZZ_2$ satisfying 
$$
\varphi(x)  +\varphi(y)  +\varphi(z)  =0\mod 2
$$
for  $x,y,z\in \partial\sigma_i$ where $i=1,2,\cdots,N$.
}
\Proof{See \cite[Proposition 3.13]{okay2022simplicial}.}
}

\subsection{Simplicial distributions}
\label{sec:Simplicial distributions}

We will write $D(U)$ for the set of probability distributions on a set $U$. It consists of functions $p:U\to \RR_{\geq 0}$ with finite support, i.e., $|\set{u\in U:\, p(u)>0}|<\infty$, such that
$$ 
\sum_{u\in U} p(u)=1.
$$
Given a function $f:U\to V$ and a distribution $p\in D(U)$ we define $f_*(p)\in D(V)$ by the formula
$$
f_*(p)(v) = \sum_{u\in f^{-1}(v)} p(u).
$$
{We will write $\delta^u$ for the delta-distribution peaked at an element $u\in U$.}
{G}iven a simplicial set $Y$ we can construct a new simplicial set $D(Y)$ whose set of $n$-simplices consists of $D(Y_n)$. The face and the degeneracy maps of this simplicial set are given by $d_i(p)= (d_i)_*(p)$ and $s_j(p)=(s_j)_*(p)$.
{We will write $\delta:Y\to D(Y)$ for the simplicial set map that sends a simplex to the delta-distribution peaked at that simplex.}

Our interest is {in} the space $D(N\ZZ_2)$. The set of $n$-simplices consists of distributions of the form 
$$
p:\ZZ_2^n \to \RR_{\geq 0}.
$$
We write $p^{a_1\cdots a_n}$ for the value $p(a_1,\cdots,a_n)$.
With this notation the face maps in dimension $2$ are given by
\begin{equation}\label{eq:faces of D of nerve}
(d_ip)^0 = \left\lbrace
\begin{array}{ll}
p^{00}+p^{10} & i=0 \\
p^{00}+p^{11} & i=1 \\
p^{00}+p^{01} & i=2. 
\end{array}
\right.
\end{equation}
See Figure (\ref{fig:two-simplex-distribution}).

\Def{\label{def:simplicial distribution}
{\rm
{A (simplicial) scenario consists of a pair $(X,Y)$ of simplicial sets representing the space of measurements and outcomes, respectively.}
A simplicial distribution on the scenario $(X,Y)$ is a simplicial set map
$$
p:X\to D(Y)
$$
We write $\sDist(X,Y)$ for the set of simplicial distributions.
}
}

When $Y=N\ZZ_2$ we will simplify the notation and write $\sDist(X)$ for the set of simplicial distributions on $X$.

\Pro{\label{pro:polytope of simplicial distributions}
Let $X$ be a simplicial set with generating $2$-simplices $\sigma_1,\sigma_2,\cdots,\sigma_N$.
Then a simplicial distribution $p:X\to D(N\ZZ_2)$ is given by a collection of distributions $p_{\sigma_k}\in D(\ZZ_2^2)$ satisfying
\begin{equation}\label{eq:di=dj}
d_i p_{\sigma_k} = d_j p_{\sigma_l}
\end{equation}
whenever $d_i\sigma_k = d_j\sigma_l$. 
In particular, $\sDist(X)$ is a (convex) polytope {with finitely many vertices.}
}
\Proof{{A simplicial set map $p:X\to D(N\ZZ_2)$ is determined by its values, i.e., the distributions $p_{\sigma_i}\in D(\ZZ_2^2)$, on the generating simplices. Note that on the remaining simplices the image is determined by the simplicial structure maps. The only relations imposed on $p_{\sigma_i}$'s come from the face maps given in Equation (\ref{eq:di=dj}).}
{Therefore} $\sDist(X)$ is the subspace obtained by intersecting $[0,1]^{4N}\subset \RR^{4N}$ with the linear equations corresponding to normalization and identifications under the face maps. This subspace is convex and bounded. Since the set of equations involved is finite there are finitely many extreme points (vertices).
}

\Ex{\label{ex:triangle simplicial distributions}
{\rm
The simplest case is when $X=\Delta^2$:
$$
\sDist(\Delta^2) = \set{(p^{00},p^{01},p^{10},p^{11})\in [0,1]^4:\, p^{00}+p^{01}+p^{10}+p^{11}=1},  
$$
{which defines a polytope in $\RR^3$, e.g., by retaining the coordinates $(p^{01},p^{10},p^{11})$.}
}}

In general, $X$ is obtained by gluing $N$ of the $\Delta^2$'s and $\sDist(X)$ is a convex polytope in $\RR^{3N}$ cut out by the linear equations (\ref{eq:di=dj}).

\Def{\label{def:contextuality}
{\rm
A simplicial distribution is called a deterministic distribution if it is of the form
$
\delta^s: X \xrightarrow{s} Y \xrightarrow{\delta} D(Y)
$
where $s:X\to Y$ is a simplicial set map. 
We will write $\dDist(X,Y)$ for the set of deterministic distributions.
There is a natural map
$$
\Theta: D(\dDist(X,Y)) \to \sDist(X,Y)
$$
defined by sending $d=\sum_s d(s) \delta^s$ to the simplicial distribution $p$ given by
$$
p_\sigma(\theta) = \sum_{s:s_\sigma=\theta} d(s). 
$$
A simplicial distribution $p$ is non-contextual if $p$ is in the image of $\Theta$. Otherwise, $p$ is called contextual. 
}
}

When $Y=N\ZZ_2$ we will simply write $\dDist(X)$ for the set of deterministic distributions on $X$.

\Ex{{\rm
Let us consider $X=\Delta^2$. 
{By Proposition \ref{pro:map to nerve} w}e have
$$
\dDist(\Delta^2) \xrightarrow{\cong} \ZZ_2^2
$$
obtained by sending $s:\Delta^2\to N\ZZ_2$ to the pair $(s_{d_2\sigma},s_{d_0\sigma})$. The corresponding determistic distribution is given by
$$
(\delta^s_\sigma)^{ab} = 
\left\lbrace
\begin{array}{ll}
1  &    (a,b) = (s_{d_2\sigma},s_{d_0\sigma}) \\
0 & \text{otherwise.}
\end{array}
\right. 
$$
For notational convenience we will write $\delta^{ab}$ when $s$ is specified by $(a,b)$ under the bijection above. 
Observe that any simplicial distribution $p$ on the triangle can be written as
$$
p = \sum_{a,b} p^{ab} \delta^{ab}.
$$
Therefore every $p$ {on $\Delta^2$} is non-contextual according to Definition \ref{def:contextuality}.
}}

\begin{ex}\label{ex:edges glued}
{\rm 
 The simplest contextual example can be obtained by identifying any two edges of the triangle. 
Formally this is expressed as a push-out diagram
$$
\begin{tikzcd}[column sep = huge, row sep =large]
\Delta^1 \sqcup \Delta^1 \arrow[d] \arrow[r,hook] & \Delta^2 \arrow[d] \\
\Delta^1 \arrow[r] & \Delta^2/x\sim z
\end{tikzcd}
$$
{where the top horizontal map sends the generating simplices of $\Delta^1$'s to {$x=d_2\sigma$} and {$z=d_1\sigma$}.}
Then we have
$$
\Theta: D(\set{ \delta^{00},\delta^{10}}) \to \set{(p^{00},p^{01}, p^{10},p^{11} ) \in [0,1]^4:\, \sum_{a,b}p^{ab}=1,\; p^{01}=p^{11} }
$$
and a  simplicial distribution $p$ is contextual if and only if $p^{11}>0$. The analysis is similar for the case where the other pairs of edges are identified. 

We can also identify all three edges to obtain $\Delta^2/x\sim y\sim z$. Then 
$$
\Theta: D(\set{ \delta^{00}}) \to \set{(p^{00},p^{01}, p^{10},p^{11} ) \in [0,1]^4:\, \sum_{a,b}p^{ab}=1,\; p^{01}= p^{10}=p^{11} }
$$
and again $p$ is contextual if and only if $p^{11}>0$. 
}
\end{ex}

\begin{ex}\label{ex:collapsing edges}
{\rm
Another topological operation we can perform on a single triangle is to collapse an edge. This is given by a push-out diagram of the form 
$$
\begin{tikzcd}[column sep = huge, row sep =large]
\Delta^1   \arrow[d] \arrow[r,hook] & \Delta^2 \arrow[d] \\
\Delta[0] \arrow[r] & \Delta^2/z\sim \ast
\end{tikzcd}
$$
Then we have
$$
\Theta: D(\set{ \delta^{00},\delta^{11}}) \xrightarrow{\cong} \set{(p^{00},0, 0,p^{11} ) \in [0,1]^4:\, p^{00}+p^{11}=1}.
$$ 
In this case every $p$ is non-contextual. Collapsing two of the edges to obtain $\Delta^2/y\sim z\sim \ast$ we {obtain}
$$
\Theta:  D(\set{ \delta^{00}}) \xrightarrow{\cong} \set{(1,0,0,0 )}
$$
and {again} every simplicial distribution is non-contextual.
The situation is exactly the same when all three edges are {collapsed}.
}
\end{ex}
 
\subsection{Twisted distributions} 
\label{sec:Twisted distributions} 

We will use the more general notion of simplicial distributions introduced in \cite{barbosa2023bundle}.

\begin{defn}
{\rm
A (simplicial) bundle scenario is a simplicial set map $\pi:E\to X$. 
A simplicial distribution on the bundle scenario $f$ is a simplicial set map $p:X\to D(E)$ that makes the following diagram commute
\begin{equation}\label{dia:simplicial distribution on pi}
\begin{tikzcd}[column sep = huge, row sep = large]
& D(E) \arrow[d,"D(\pi)"] \\
X \arrow[ru,"p"] \arrow[r,"\delta"] & D(X)
\end{tikzcd}
\end{equation}
A deterministic distribution on $f$ is a simplicial set map of the form $\delta^s:X\xrightarrow{s} E \xrightarrow{\delta} D(E)$ where $s:X\to E$ is a section of $\pi$.
}
\end{defn}

The earlier notion given in Definition \ref{def:simplicial distribution} can be recovered by   considering simplicial distributions 
on the bundle scenario given by the projection map
$$
\pi: Y\times X\to X
$$
For twisted distributions we will need the notion of twisted products.
Twisted products model principal bundles \cite{May67}. {Let us write $H_n(X)$ and $H^n(X)$ for the $n$-th homology and cohomology groups with coefficients in $\ZZ_2$, respectively.} When $X$ is $2$-dimensional there is a well-known classification theorem: the set of isomorphism classes of principal bundles with fiber $N\ZZ_2$ is in bijective correspondence with the classes in the second cohomology group $H^2(X)$.
Let us recall the definition of the $n$-th cohomology group of a simplicial set \cite{goerss2009simplicial}. {Given a simplicial set $X$ let us write $C^n(X)$ for the set $\ZZ_2^{X_n}$ of functions $\alpha:X_n\to \ZZ_2$. The coboundary map $\delta_n:C^n(X)\to C^{n+1}(X)$ is defined by
$$
\delta_n \alpha(x) = \sum_{i=0}^n (-1)^{i} \alpha(d_ix).
$$
Then $H^n(X)$ is defined as the quotient of the kernel of $\delta_n$ {by} the image of $\delta_{n-1}$. A function $\alpha:X_n\to \ZZ_2$ is called a cocycle if it belongs to the kernel of $\delta_n$. It is called normalized if $\alpha(x)=0$ for every degenerate $n$-simplex $x$.}

\Def{{\rm
Let $\beta:X_2\to \ZZ_2$ be a normalized $2$-cocycle.
The twisted product $N\ZZ_2\times_\beta X$ is the simplicial set whose set of $n$-simplices is given by $(N\ZZ_2)_n\times X_n$. The face map in dimension $2$ is given by
\begin{equation}\label{eq:face of twisted}
d_i((a,b),\sigma) =
\left\lbrace 
\begin{array}{ll}
(b+\beta(\sigma),d_0\sigma) & i=0 \\
(a+b,d_1\sigma) & i=1\\
(a,d_2\sigma) & i=2.
\end{array}
\right.
\end{equation}
}}

{Note that the $d_0$-face is twisted by the cocycle $\beta$.}
Projecting onto the second coordinate gives a principal $N\ZZ_2$-bundle
$$
\pi: N\ZZ_2\times_\beta X \to X
$$
We are interested in simplicial distributions on such maps.

\Def{\label{def:twisted distribution}
{\rm
A $\beta$-twisted distribution is a simplicial distribution on $\pi:N\ZZ_2\times_\beta X\to X$. A $\beta$-twisted deterministic distribution is a deterministic distribution on $\pi$.
We will write $\sDist_\beta(X)$ and $\dDist_\beta(X)$ for the sets of $\beta$-twisted simplicial and deterministic distributions on $X$, respectively.
}
}

{Let us unravel the definition of a twisted distribution when $X$ is $2$-dimensional generated by $\sigma_1,\cdots,\sigma_N$. A twisted simplicial distribution associates a distribution $p_{\sigma_i}\in D(\ZZ_2^2\times X_2)$ with each $2$-simplex $\sigma_i$. The commutativity of diagram   (\ref{dia:simplicial distribution on pi}) implies that the support of the distribution is contained in $\ZZ_2^2 \times \set{\sigma_i}$. 
Face maps can be worked out from Equation (\ref{eq:face of twisted}). Only the the $d_0$-face is twisted
\begin{equation}\label{eq:d0 face}
(d_0p_\sigma)^0 = \left\lbrace
\begin{array}{ll}
p^{00}+p^{10} & \beta(\sigma)=0\\
p^{11}+p^{01} & \beta(\sigma)=1.
\end{array}
\right.
\end{equation}
{See Figure (\ref{fig:two-simplex-twisted-distribution}).}

A version of Proposition \ref{pro:polytope of simplicial distributions} holds for $\sDist_\beta(X)$ once the $d_0$-face is twisted in this way.}
{We will provide a more explicit description of the resulting polytope.}
We define the correlation function
\begin{equation}\label{dia:correlation function}
c: X_1 \to \RR
\end{equation} 
by $c(x) = p_x^0 - p_x^1$. The probabilities $p_\sigma^{ab}$ for each $2$-simplex $\sigma\in X_2$ can be recovered from the correlation function. 

\Lem{\label{lem:probabilities from correlations}
We have  
\begin{equation}\label{eq:probabilities from correlations}
p_\sigma^{ab} = \frac{1}{4}(1+ (-1)^a c(x_2) + (-1)^{b+\beta(\sigma)} c(x_0) + (-1)^{a+b} c(x_1)).
\end{equation}
}
\Proof{
This formula can be verified using the marginal relations (e.g., $d_0$ face as in Equation (\ref{eq:d0 face}) and others) and the relation $p_{x_i}^1 = 1-p_{x_i}^0$. For example, for $(a,b)=(0,0)$ and $\beta(\sigma)=1$ we have
$$
\begin{aligned}
\frac{1}{4}(1+ c(x_2) - c(x_0) +  c(x_1)) &= \frac{1}{4}(1+ (2p_{x_2}-1) - (2p_{x_0}-1) +  (2p_{x_1}-1)) \\
&= \frac{1}{4}(1+ (2(p^{00}_\sigma+p^{01}_\sigma)-1) - (2(p^{11}_\sigma+p^{01}_\sigma)-1) +  (2(p^{00}_\sigma+p^{11}_\sigma)-1))\\
&=p^{00}_\sigma.
\end{aligned}
$$
}

Using this result we can provide a more explicit description for the polytope of twisted simplicial distributions. For a $m\times d$ matrix $M$ and a column vector $b$ of size $m$ we will write 
$$P(M,b)=\set{t\in \RR^d:\, Mt \geq b} $$ 
for the corresponding polytope in $\RR^d$.
Let $\one$ denote a column vector consisting of $1$'s. 

\Pro{\label{pro:polytope of twisted simplicial distributions}
Let $d=|X_1|$ and $m=|X_2\times \ZZ_2^2|$. Then
$$
\sDist_\beta(X) = P(M,-\one)
$$
where $M$ is the $m\times d$ matrix defined by
$$
M_{(\sigma,ab),x} = \left\lbrace
\begin{array}{ll}
(-1)^{b+\beta(\sigma)} & x=d_0\sigma \\
(-1)^{a+b} & x=d_1\sigma\\ 
(-1)^a & x=d_2\sigma.
\end{array}
\right.
$$ 
}
\Proof{
This follows directly from Lemma \ref{lem:probabilities from correlations}.
}

\Pro{\label{pro:properties}
We have the following properties.
\begin{enumerate}
\item The set $\dDist_\beta(X)$ is non-empty if and only if $[\beta]=0$.
\item There is a bijection
$$
\sDist_\beta(X) \xrightarrow{\cong} \sDist_\alpha(X)
$$
when $[\beta]=[\alpha]$.
\end{enumerate}
}
\begin{proof}
{Part (1): The cohomology class $[\beta]$ vanishes if and only if the principal bundle $N\ZZ_2\times_\beta X\to X$ trivial. The latter holds if and only if it admits a section. Part (2): The cohomology classes coincide if and only the corresponding principal bundles are isomorphic {(as principal bundles)}. This gives a commutative diagram where the top arrow is an isomorphism}
$$
\begin{tikzcd}[column sep = huge, row sep = large]
D(N\ZZ_2\times_\beta X) \arrow[rr,"\cong"] \arrow[dr] && D(N\ZZ_2\times_\alpha X) \arrow[dl] \\
&D(X)&  
\end{tikzcd}
$$
{Therefore the resulting sets of simplicial distributions are in bijective correspondence.}
\end{proof}

\begin{figure}[h!] 
  \centering
  \includegraphics[width=.5\linewidth]{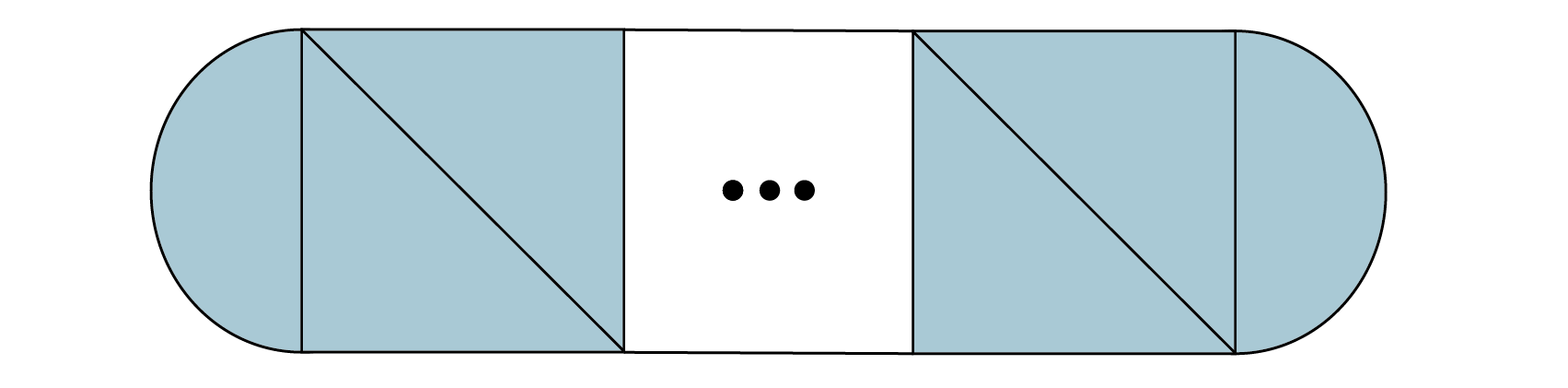}
\caption{
}
\label{fig:line-N}
\end{figure}

\Ex{\label{ex:line}
{\rm
Let $D_N$ be a simplicial set obtained by gluing $N$ many $2$-simplices to obtain a disk, e.g., as in Figure (\ref{fig:line-N}). {For a more general definition see \cite[Definition 3.1]{kharoof2023topological}.} Since the resulting space is contractible we have $H_2(D_N)=0$. Therefore for any normalized cocycle $\beta:(D_N)_2\to \ZZ_2$ we have $[\beta]=0$. By Proposition \ref{pro:properties} we have $\dDist_\beta(D_N)\neq \emptyset$ and 
$$
\sDist_\beta(D_N) \cong \sDist(D_N).
$$
Then the gluing lemma of \cite{okay2022simplicial} implies that every $p\in \sDist_\beta(D_N)$ is non-contextual. 
}}

For a simplicial subset $Z\subset X$ and a simplicial distribution on $X$ we write $p|_Z$ for the composite $Z \xrightarrow{i} X \xrightarrow{p} D(E)$ where $i$ is the inclusion map. 
{The following simple observation will be very useful in analyzing the vertices of the polytope of twisted distributions.}
 
\Lem{(\!\cite{okay2022mermin})
\label{lem:two implies three}
Let $Z\subset X$ be a simplicial subset with a single generating $2$-simplex $\sigma$. If  two of the faces, say $x,y\in \partial\sigma$, satisfy $p_x=\delta^a$ and $p_y=\delta^b$ for some $a,b\in \ZZ_2$ then $p|_Z$ is a 
deterministic distribution.
}
\Proof{
{When the distribution is deterministic on two of the edges this {forces} three of the four parameters in $\sDist_\beta(\Delta^2)\subset [0,1]^4$ to be zero. Together with the normalization condition we obtain a unique solution, i.e., a deterministic distribution.
}
}

\begin{figure}[h!]
\centering
\begin{subfigure}{.49\textwidth}
  \centering
  \includegraphics[width=.4\linewidth]{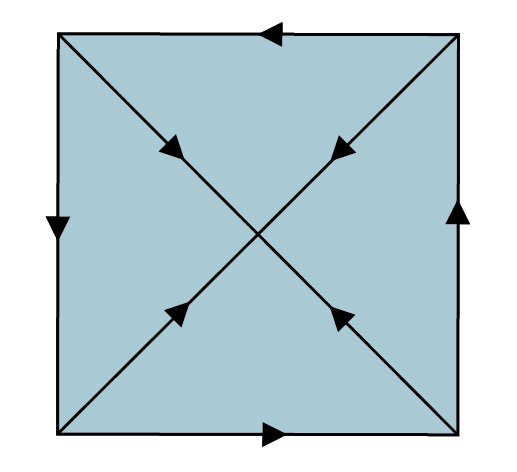}
  \caption{}
  \label{fig:CHSH}
\end{subfigure}%
\begin{subfigure}{.49\textwidth}
  \centering
  \includegraphics[width=.4\linewidth]{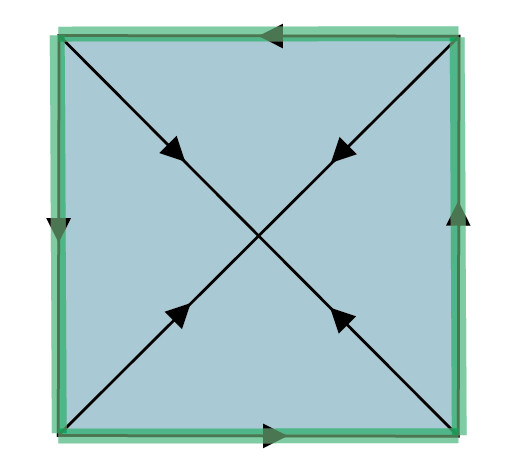}
  \caption{}
  \label{fig:PR}
\end{subfigure}
\caption{{(a) The measurement space of the CHSH scenario. (b) PR box with $a_1+a_2+a_3+a_4=1\mod 2$. Green color indicates the deterministic edges.}
}
\label{fig:CHSH-PR}
\end{figure}

\Ex{\label{ex:CHSH}
{\rm {
The measurement space $\tilde C_4$ of the Clauser--Horne--Shimony--Holt (CHSH) scenario  consists of four triangles $\sigma_1,\sigma_2,\sigma_3,\sigma_4$ glued as in Figure (\ref{fig:CHSH}). This is a special case of   cycle scenarios we will examine in Section \ref{sec:N-cycle-scenario}. It is well-known that the polytope $\sDist(\tilde C_4)$ has two kinds of vertices consisting of the deterministic distributions and contextual distributions known as the Popescu--Rohrlich (PR) boxes \cite{popescu1994quantum}; see also \cite[Example 3]{kharoof2023topological}. PR boxes are characterized by the property that the restriction $p|_{\partial\tilde C_4}$ to the boundary, which consists of four edges $x_1,x_2,x_3,x_4$, is a deterministic distribution $\delta^s$ such that
$$
\sum_{i=1}^4 a_i = 1 \mod 2
$$
where $\delta^s_{x_i}=\delta^{a_i}$ for $i=1,2,\cdots,N$.}
}}

\subsection{Product of distributions}
\label{sec:Product of distributions}


{In \cite{kharoof2022simplicial} a product is introduced for simplicial distributions. We can  extend this product to $2$-dimensional twisted distributions.}
The convolution product of $p,q\in D(\ZZ_2^n)$ is the distribution $p\ast q$ defined by
$$
p\ast q(c) = \sum_{a+b=c} p(a)q(b)
$$
where the sum runs over $a,b\in \ZZ_2^n$ such that $a+b=c$.
Using the convolution product we define a map
$$
\sDist_\alpha(X) \times \sDist_\beta(X) \to \sDist_{\alpha+\beta}(X)
$$
by sending $(p,q)$ to the distribution $p\cdot q$ given by
$$
(p\cdot q)_\sigma = p_\sigma \ast q_\sigma.
$$

\Lem{\label{lem:product-twisted}
The product $p\cdot q$ is an $(\alpha+\beta)$-twisted distribution.
}
\Proof{
We have
$$
\begin{aligned}
((p\cdot q)_{d_0\sigma})^0 &=  p_{d_0\sigma}^0 q_{d_0\sigma}^0 +  p_{d_0\sigma}^1 q_{d_0\sigma}^1  \\
&= \sum_a p_\sigma^{a(\alpha(\sigma))}  \sum_b q_\sigma^{b(\beta(\sigma))} + \sum_a p_\sigma^{a(\alpha(\sigma)+1)}  \sum_b q_\sigma^{b(\beta(\sigma)+1)} \\
&= (p\cdot q)^{0(\alpha(\sigma)+\beta(\sigma))} +(p\cdot q)^{1(\alpha(\sigma)+\beta(\sigma))}\\
&=(d_0 (p\cdot q)_\sigma )^0.
\end{aligned}
$$
{Similarly one can verify that $p\cdot q$ is compatible with the remaining simplicial structure maps. Commutativity of diagram (\ref{dia:simplicial distribution on pi}) follows from the observation that the support of $p\cdot q$ is contained in $\ZZ_2^2\times \set{\sigma_i}$ by definition of the product.}
}

  
It is instructive to consider the action of $\dDist_\alpha(X)$ on $\sDist_\beta(X)$ induced by this product. Note that because of part (1) of Proposition \ref{pro:properties} we assume $[\alpha]=0$, or for computational simplicity $\alpha=0$.  Then  we have
\begin{equation}\label{eq:product of deterministics}
(\delta^{ab}\cdot q)^{cd}_\sigma = q_\sigma^{(c+a)(b+d)}
\end{equation}
where $\delta^{ab}\in \dDist(X)$ and $q\in \sDist_\beta(X)$.

\subsection{Cohomology exact sequence}
\label{sec:Cohomology exact sequence}

Let $Z\subset X$ be a simplicial subset. 
Given a normalized $2$-cocycle $\beta:X_2\to \ZZ_2$  we will write $\beta|_Z$ for
 the
pull-back $i^*\beta$ along the inclusion map $i:Z\to X$.
We assume   $[\beta|_Z]=0$ so that there exists a normalized $1$-cochain $s:Z_1\to \ZZ_2$ such that $\beta|_Z = \delta s$. 
We will write 
$$
\sDist_\beta(X,s)=\set{p\in \sDist_\beta(X):\, p|_Z=\delta^s}.
$$ 
Consider the cofiber sequence
$$
Z\xrightarrow{i} X\xrightarrow{q} X/Z
$$
and the  associated cohomology long exact sequence 
$$
H^1(X/Z) \to H^1(X) \to H^1(Z) \xrightarrow{\zeta} H^2(X/Z)
$$
Let $\zeta(s)$ denote the $2$-cocycle obtained by the snake lemma \cite{weibel1995introduction}, i.e., first extending $s$ to $X$ and applying the coboundary:
\begin{itemize}
\item Let $\tilde s:X_1\to \ZZ_d$ denote the $2$-cochain defined by
$$
\tilde s(x) =
\left\lbrace
\begin{array}{ll}
s(x) & x\in Z_1 \\
0    & x\in X_1-Z_1.
\end{array}
\right.
$$
\item By applying the coboundary map $\delta:C_1(X)\to C_2(X)$ we obtain
$$
\delta\tilde s(\sigma) =  \tilde s(d_0\sigma)-\tilde s(d_1\sigma)+\tilde s(d_2\sigma)
$$
where $\sigma \in X_2$.

\item Since $\delta\tilde s|_Z=0$ it comes from a $2$-cochain, which will be denoted by $\zeta(s):(X/Z)_2\to \ZZ_2$.
\end{itemize}
{See also \cite[Section 5]{okay2022simplicial}} 
 
\Lem{\label{lem:delta-tilde-s}
The deterministic distribution $\delta^{\tilde s}$ on $X$ defined by
$$
(\delta^{\tilde s}_\sigma)^{ab} = \left\lbrace
\begin{array}{ll}
1 & (a,b)=(\tilde s(d_2\sigma),\tilde s(d_0\sigma))\\
0 &\text{otherwise}
\end{array}
\right.
$$
is a $\delta\tilde s$-twisted distribution.} 
\Proof{{This is direct verification.}}

\begin{pro}\label{pro:zeta-twisted}
We have a commutative diagram
$$
\begin{tikzcd}
D(\dDist_\beta(X,s)) \arrow[r,"\Theta"] \arrow[d,"D(\phi)","\cong"'] &  \sDist_\beta(X,s) \arrow[d,"\phi","\cong"'] \\
D(\dDist_{\beta+\zeta(s)}(X/Z)) \arrow[r,"\Theta"] & \sDist_{\beta+\zeta(s)}(X/Z)
\end{tikzcd}
$$
where $\phi(p)=\delta^{\tilde s}\cdot p$. Both vertical arrows are isomorphisms. 
\end{pro}
\Proof{
Let $p\in \sDist_\beta(X)$ be such that $p|_Z = \delta^s$ for some $s$ satisfying $\delta s=\beta|_Z$.
By Lemma \ref{lem:product-twisted} and \ref{lem:delta-tilde-s} the distribution $\delta^{\tilde s}\cdot p$ is $(\beta+\zeta(s))$-twisted {and
$$
(\delta^{\tilde s}\cdot p)|_Z = \delta^{\tilde s}|_Z\cdot p|_Z = \delta^s \cdot \delta^s = \delta^0
$$
where $\delta^0:Z\to D(N\ZZ_2)$ is the constant map (with image the unique vertex). Note that in the last equation we used Equation (\ref{eq:product of deterministics}). Therefore $\delta^{\tilde s} \cdot p$ factors through the quotient $X/Z$.
} 
The inverse of $\phi$ is given by the composite
$$
\phi^{-1}:\sDist_{ \beta+ \zeta(s)}(X/Z) \xrightarrow{q^*} \sDist_{ \beta+ d\tilde s}(X) \xrightarrow{\delta^{\tilde s}\cdot} \sDist_{ \beta}(X)
$$
The maps $\phi$ and $\phi^{-1}$ restrict to a bijection between $\dDist_\beta(X,s)$ and $\dDist_{\beta+\zeta(s)}(X/Z)$.
}

{ 
\Cor{\label{cor:quotient non-contextual}
If $\phi(p)$ is non-contextual then $p$ is non-contextual.
}
}

\begin{figure}[h!]
\centering
\begin{subfigure}{.49\textwidth}
  \centering
  \includegraphics[width=.4\linewidth]{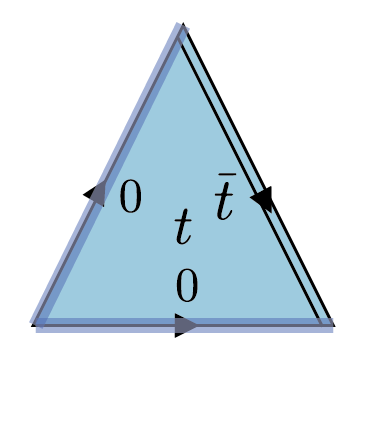}
  \caption{}
  \label{fig:identified-0}
\end{subfigure}%
\begin{subfigure}{.49\textwidth}
  \centering
  \includegraphics[width=.4\linewidth]{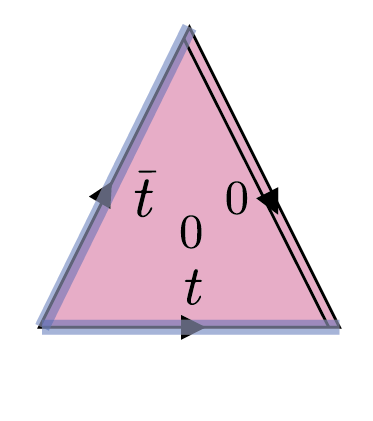}
  \caption{}
  \label{fig:identified-1}
\end{subfigure}
\caption{{Twisted simplicial distributions on a triangle where $x$ is identified with $z$ and $y$ is collapsed: (a) $\beta(\sigma)=0$ and (b) $\beta(\sigma)=1$. The resulting polytope can be identified with the unit interval $[0,1]$. Here $\bar t = 1-t$.}
}
\label{fig:x and z identified y collapsed}
\end{figure}

\Ex{\label{ex:x and z identified y collapsed}
Let us consider $\sDist(X)$ where $X=\Delta^2/x\sim z$; see Example \ref{ex:edges glued}.
We take $Z$ to be the simplicial subset given by the edge $y$. If the restriction $p|_Z$  is deterministic then it is either $\delta^0$ or $\delta^1$. 
Then by Proposition \ref{pro:zeta-twisted} we have an isomorphism
$$
\sDist(X,s) \xrightarrow{\cong} \sDist_\beta(X/y\sim \ast)
$$ 
Note that $\beta(\sigma)=a$ if $p_y^a=\delta^a$. See Figure (\ref{fig:x and z identified y collapsed}).
}

\section{Distributions on graphs}
\label{sec:Distributions on graphs} 
\begin{figure}[h!]
\centering 
  \includegraphics[width=.6\linewidth]{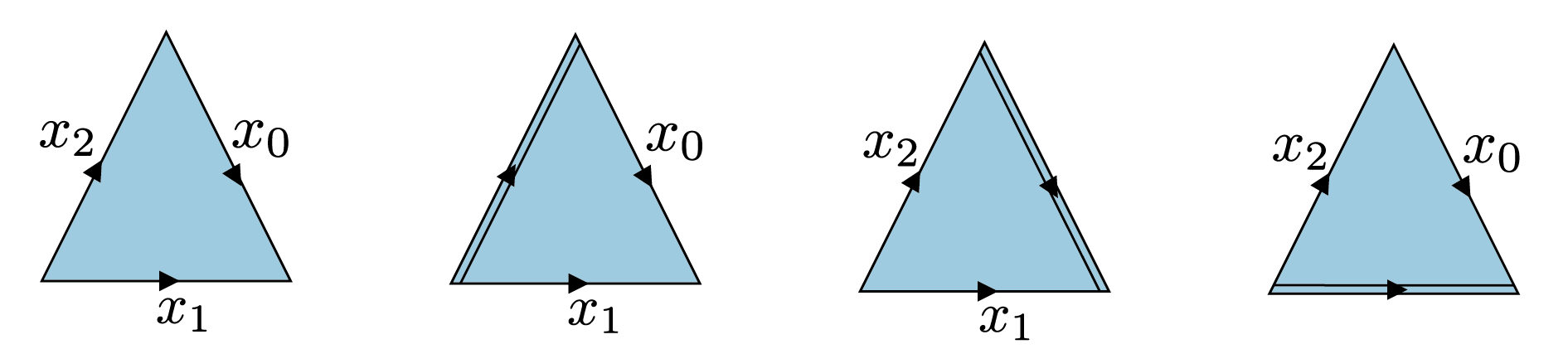}
  \caption{} 
\label{fig:want}
\end{figure}

Our goal in this section is to describe twisted distributions as distributions on graphs. 
This is achieved by associating a graph with a simplicial set.
Throughout the paper we only consider simple graphs, i.e., graphs with no loops and parallel edges. 
To land in simple graphs we impose some conditions on the simplicial sets that represent the measurements. 
We will restrict to a measurement space $X$ given by   a simplicial set    
\begin{itemize}
\item generated by the $2$-simplices $\sigma_1,\sigma_2,\cdots,\sigma_N$, and
\item each set 
{$\partial\sigma_k$}
of 
edges in the boundary consists of  either three distinct {non-degenerate} edges $\set{x_0,x_1,x_2}$ or two distinct {non-degenerate} edges $\set{x_i,x_j}$, where $i>j$, {and a remaining degenerate edge}.
\end{itemize}
Therefore we have the situation depicted in Figure (\ref{fig:want}).

\subsection{Distributions on graphs}

We begin by constructing a bipartite graph associated to the simplicial set $X$. 
A bipartite graph $\Gamma$ consists of a vertex set $V(\Gamma)$, partitioned into two sets $V^0(\Gamma)\sqcup V^1(\Gamma)$, and an edge set $E(\Gamma)$ connecting the vertices from these two sets. {We will also consider graphs with a sign given by a function $\gamma:E(\Gamma)\to \set{\pm 1}$. 

\Def{\label{def:graph of X}{\rm
{Given a finite $2$-dimensional simplicial set $X$}
let $\Gamma(X)$ denote the bipartite graph with
\begin{itemize}
\item vertex set $V=X_1^\circ \sqcup X_2^\circ$,  
\item edge set $E$ consisting of $\set{x,\sigma}$ where $x\in \partial\sigma^\circ$.
\end{itemize} 
}}

Next, we enlarge this graph by including the set of outcomes into the picture. The idea is to replace each vertex corresponding to $\sigma$ by four vertices labeled by $s_\sigma^{ab}$ where $a,b\in \ZZ_2$. The new vertices are also connected to the same vertices corresponding to the $1$-simplices in $\partial\sigma^\circ$. These edges of the graph are also assigned a sign $\pm 1$ indicating the outcomes. See Figure (\ref{fig:local}) for the local picture over a triangle with two kinds of boundaries, one with $|\partial\sigma^\circ|=3$ and $2$.

\Def{\label{def:enlarged graph of X}
{\rm
Given a normalized $2$-cocycle $\beta:X_2\to \ZZ_2$ we define a signed bipartite graph $\Gamma_\beta(X)$ with
\begin{itemize}
\item vertex set $X_1^\circ \sqcup (X_2^\circ\times \ZZ_2^2)$,
\item edge set   consisting of $\set{x,s_\sigma^{ab}}$ {where $x\in \partial\sigma^\circ$,}
\item and sign given by
\begin{equation}\label{eq:sign gamma}
\gamma(x,s_\sigma^{ab})=(-1)^{s_\sigma(x)}
\end{equation}
where $s_\sigma:\partial\sigma\to \ZZ_2$ is defined by 
$$
s_\sigma(x) = \left\lbrace
\begin{array}{ll}
(-1)^{b+\beta(\sigma)} & x=d_0\sigma \\
(-1)^{a+b} & x=d_1\sigma \\
(-1)^{a} & x=d_2\sigma.
\end{array}
\right.
$$
\end{itemize}   
}}

Next, we introduce the notion of a distribution on the bipartite graph associated to the pair $(X,\beta)$. The goal is to capture the notion of twisted simplicial distributions as distributions on graphs. {The next definition is motivated by Proposition \ref{pro:polytope of twisted simplicial distributions}.}

\Def{\label{def:distribution on graph}
{\rm
A  
distribution on $\Gamma_\beta(X)$ is a function $p: X_2^\circ\times \ZZ_2^2\to \RR_{\geq 0}$ such that
$$
\sum_{a,b} p(s_\sigma^{ab}) =1 
$$
for every $\sigma\in X_2^\circ$ and 
\begin{equation}\label{eq:match}
\sum_{a,b} \gamma(x,s_\sigma^{ab}) p(s_\sigma^{ab}) = \sum_{a,b} \gamma(x,s_{\sigma'}^{ab}) p(s_{\sigma'}^{ab})
\end{equation}
for every $\sigma,\sigma'\in X_2^\circ$ such that $x\in \partial\sigma^\circ\cap\partial\sigma'^\circ$.
In addition, for simplices $\sigma\in X_2^\circ$ whose boundary contains a single degenerate edge $x'$ we also require that {
\begin{equation}\label{eq:last condition}
\left\lbrace
\begin{array}{ll}
p(s_\sigma^{(0+\beta(\sigma))(0+\beta(\sigma))})+p(s_\sigma^{(1+\beta(\sigma))(0+\beta(\sigma))})  =1 &  x'=d_0\sigma \\
p(s_\sigma^{00})+p(s_\sigma^{11})=1    &  x'=d_1\sigma \\
p(s_\sigma^{00})+p(s_\sigma^{01}) =1  &  x'=d_2\sigma .
\end{array}
\right. 
\end{equation}
We write $\Dist(\Gamma_\beta(X))$ for the set of  
distributions on the graph $\Gamma_\beta(X)$.}
}
}

\begin{figure}[h!]
\centering
\begin{subfigure}{.49\textwidth}
  \centering
  \includegraphics[width=.4\linewidth]{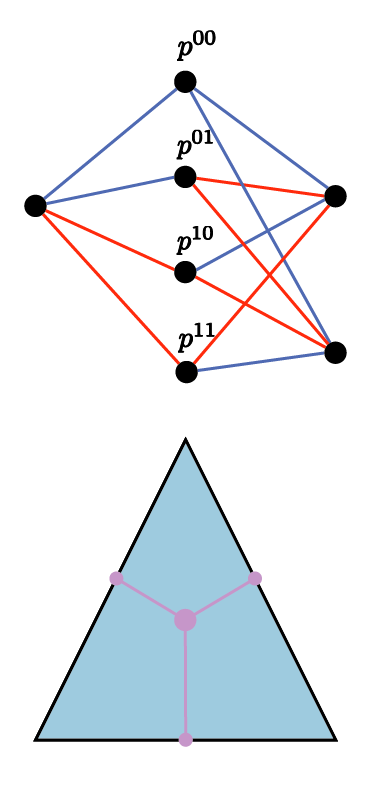}
  \caption{}
  \label{fig:want-a}
\end{subfigure}%
\begin{subfigure}{.49\textwidth}
  \centering
  \includegraphics[width=.4\linewidth]{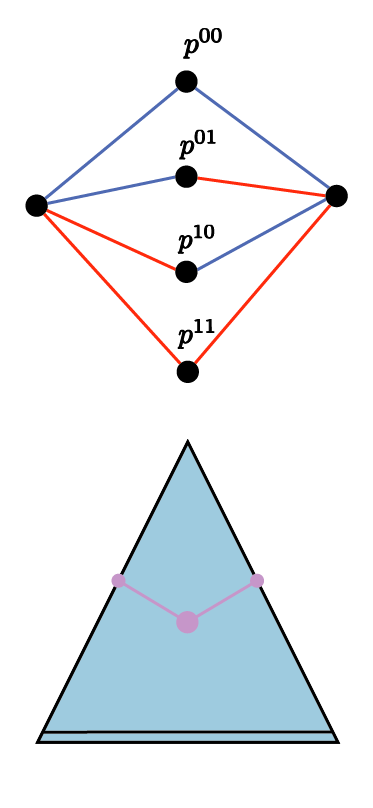}
  \caption{}
  \label{fig:want-b}
\end{subfigure}
\caption{Signs on the edges induced by $s_\sigma^{ab}$ are indicated by blue and red color for $+1$ and $-1$, respectively. The probabilities are given by $p^{ab}=p(s_\sigma^{ab})$.  In this example $\beta(\sigma)=0$.
}
\label{fig:local}
\end{figure}

Equation (\ref{eq:last condition}) has the following interpretation. As in simplicial distributions we want $p_{x'}=\delta^0$ for each degenerate edge $x'$. This condition is imposed via this equation separately  since the degenerate edges of $X$ does not appear in the graph $\Gamma_\beta(X)$.
See Figure (\ref{fig:local}) for the local picture over a simplex. {Definition \ref{def:distribution on graph} is tailored so that the following identification can be done.}

\Pro{\label{pro:iso between graph distributions and simplicial distributions}
Sending a twisted distribution $p\in \sDist_\beta(X)$ to the  distribution on the graph $\Gamma_\beta(X)$ defined by
$$
p(s_\sigma^{ab}) = p_\sigma^{ab}
$$
gives a bijection of convex sets
\begin{equation}\label{eq:iso-simplicial-vs-graph}
\sDist_\beta(X)\xrightarrow{\cong} \Dist(\Gamma_\beta(X))
\end{equation} 
}
\Proof{
Follows from Proposition \ref{pro:polytope of twisted simplicial distributions}.
}

{We will identify twisted simplicial distributions and   distributions on the associated enlarged graph.}

\subsection{Rank of a distribution}
\label{sec:Rank of a distribution}

\Def{\label{def:Gamma p}{\rm
Given a distribution $p\in \Dist(\Gamma_\beta(X))$ we consider the induced signed subgraph ${\Gamma_\beta(X,p)}\subset  \Gamma_\beta(X)$ determined by the vertex set
$$
X_1^\circ \sqcup \set{s_\sigma^{ab}\in  X_2^\circ\times \ZZ_2^2:\, p(s_\sigma^{ab})=0 }.
$$
The edge set  is determined by the vertices, that is, $\set{x,s_\sigma^{ab}}$ is an edge if $p(s_\sigma^{ab})=0$ and $x\in \partial\sigma^\circ$.
}}
 
Let $A({\Gamma_\beta(X,p)})$ denote the adjacency matrix of the signed graph.
Its rows and columns are indexed by the vertices of the graph and its entries are $0,\pm 1$ indicating whether there exists an edge  connecting the pair of vertices (taking into account the sign).
Since the graph is bipartite the adjacency matrix has the form 
$$
A(\Gamma_b)  = \twobytwo{\zero}{B({\Gamma_\beta(X,p)})}{B({\Gamma_\beta(X,p)})^T}{\zero}
$$
where 
$$
B({\Gamma_\beta(X,p)})_{s_\sigma^{ab},x} = \left\lbrace \begin{array}{ll}
\gamma(x,s_\sigma^{ab}) & \set{x,s_\sigma^{ab}} \in E(\Gamma_\beta(X,p)) \\
0 & \text{otherwise.}
\end{array} \right.
$$

\Def{\label{def:rank}
The rank of $p\in \Dist(\Gamma_\beta(X))$ is defined to be the rank of the matrix $B({\Gamma_\beta(X,p)})$. 
The rank of a twisted distribution $p\in \sDist_\beta(X)$ is the rank of the associated distribution on the graph.
 
} 


\Cor{\label{cor:vertex}
A twisted distribution $p\in \sDist_\beta(X)\subset {\RR^{|X_1^\circ|}}$ is a vertex if and only if $\rank(p)={|X_1^\circ|}$.
}
\Proof{This observation follows from the general theory of polytopes; see \cite[Theorem 18.1]{chvatal1983linear}.
}

Note that the polytope  $\sDist_\beta(X)$ is not full-dimensional in $\RR^{|X_1^\circ|}$ as its dimension is given by 
$$
d(X)=|X_1^\circ| - \sum_{\sigma\in X_2^\circ} |\partial \sigma-\partial\sigma^\circ |.
$$
This dimension count follows from Example \ref{ex:triangle simplicial distributions} and \ref{ex:collapsing edges}.

\Def{\label{def:deterministic-edge-triangle-Zp}
{\rm
Let $p\in \sDist_\beta(X)$ be a twisted distribution. 
A non-degenerate $1$-simplex $x$ is called a {deterministic edge} if $p_x=\delta^a$ for some 
$a\in \ZZ_2$. Similarly a non-degenerate $2$-simplex $\sigma$ is called a {deterministic triangle} if $p_\sigma=\delta^{ab}$ for some
$(a,b)\in \ZZ_2^2$.
We will write $Z_p\subset X$ for the simplicial subset generated by the deterministic edges and triangles with respect to $p$.
}}

Recall the bijection in Proposition \ref{pro:zeta-twisted}:
$$\phi:\sDist_\beta(X,s) \to \sDist_{\bar\beta}(\bar X)$$ 
where $\bar X=X/Z_p$ and $\bar\beta=\beta+\zeta(s)$.
In the next result we show that the rank of $\bar p=\phi(p)$ is related to the rank of $p$.

\Lem{\label{lem:det-rank}
For $p\in \sDist_\beta(X)$ we have 
\begin{equation}\label{eq:rank-quotient-X1}
\rank(p)= \rank({\bar p})+|(Z_p^\circ)_1|.
\end{equation}
}
\Proof{ 
For $\sigma\in X_2^\circ$ let us write ${\Gamma_\beta(X,p)}|_\sigma \subset {\Gamma_\beta(X,p)}$ for the induced subgraph on the vertices $\partial \sigma^\circ \sqcup (\set{\sigma}\times \ZZ_2^2)$.
We will take the quotient in steps, that is, one deterministic simplex at a time. 
{By Lemma \ref{lem:two implies three} each $\sigma \in (Z_p)_2^\circ$ satisfies $|\partial\sigma^\circ|=2$ or $3$.}
{Therefore we have} 
$$
X=X^{(0)} \to X^{(1)} \to \cdots \to X^{(i)} \to X^{(i+1)}\to \cdots \to X^{(t-1)} \to  X^{(t)}= X/Z_p
$$
where $X^{(i+1)}$ is obtained from $X^{(i)}$ by killing a non-degenerate deterministic edge or a non-degenerate deterministic triangle. Let us consider these two types of quotients:
\begin{enumerate}
\item Killing a non-degenerate deterministic edge $x$ which does not belong to a deterministic triangle: Let $\sigma$ be a non-degenerate triangle whose boundary contains $x$. Since $p_\sigma$ is not deterministic $|\partial \sigma^\circ|=3$ {(by Lemma \ref{lem:two implies three})}. 
We have
$$
B({\Gamma_\beta(X,p)}|_\sigma) = \begin{bmatrix}
(-1)^{s(x)} & 1 & (-1)^{s(x)+\beta(\sigma)} \\
(-1)^{s(x)} & -1 & (-1)^{s(x)+\beta(\sigma)+1}
\end{bmatrix} \sim 
\begin{bmatrix}
0 & 1 & (-1)^{s(x)+\beta(\sigma)} \\
1 & 0 & 0 
\end{bmatrix}
$$
{where $\sim$ indicates a sequence of elementary row operations.}
On the other hand, we have
$$
B({\Gamma_{\bar \beta}(\bar X,\bar p)}|_\sigma) =  
\begin{bmatrix}
1 &  (-1)^{\bar \beta(\sigma)} \\
-1 & (-1)^{\bar \beta(\sigma)+1}.
\end{bmatrix}  
=
\begin{bmatrix}
1 &  (-1)^{s(x)+ \beta(\sigma)} \\
-1 & (-1)^{s(x)+ \beta(\sigma)+1}  
\end{bmatrix}  
\sim
\begin{bmatrix}
1 &  (-1)^{s(x)+ \beta(\sigma)} \\
0 & 0  
\end{bmatrix}  
$$
where $\bar\beta(\sigma) = \beta(\sigma)+\zeta(s)(\sigma)$.
Therefore we have
$
\rank(B({\Gamma_\beta(X,p)}|_\sigma)) = \rank(B({\Gamma_{\bar \beta}(\bar X,\bar p)}|_\sigma)) +1$. That is, the rank at $\sigma$ has two parts. First part comes from the new distribution on the quotient space and the second part from the deterministic edge. Therefore Equation (\ref{eq:rank-quotient-X1}) holds.

\item Killing a non-degenerate deterministic triangle $\sigma$: Since $p_\sigma$ is deterministic $B({\Gamma_\beta(X,p)}|_\sigma)$ is a $3\times 3$-matrix of rank $3$.  
This time all the contribution comes from the deterministic edges and thus Equation (\ref{eq:rank-quotient-X1}) holds. 
\end{enumerate} 
{When a deterministic edge is killed, either as in (1) or part of a deterministic triangle as in (2), the corresponding column in $B({\Gamma_\beta(X,p)})$ can be removed. Once all such columns are removed the remaining matrix has the same rank as $B({\Gamma_{\bar\beta}(\bar X,\bar p)})$ by the local computations at (1) and (2).}
}

By this lemma we can assume that $p$ does not give rise to any deterministic simplex (edge or triangle). 
Otherwise, we can always take a quotient by the simplicial subset $Z_p$ and use the formula in Lemma \ref{lem:det-rank}.  
If $\partial\sigma$ contains a degenerate edge {$x'$} then $p(s_\sigma^{ab})=0$ for precisely two of the pairs $(a,b)\in \ZZ_2^2$. 
Note that these pairs are of the form $(a,b)$ and $( a+1, b+1)$. 
We remove one of these vertices for each such simplex keeping  only $s_\sigma^{ab}$ with $a=0$.

\Def{\label{def:signed Gamma X gamma p}
{\rm
Let $p\in \sDist_\beta(X)$ be a twisted distribution {with no deterministic simplices (edges or triangles).} 
Let ${\Gamma^0_\beta(X,p)}$ denote the induced {signed} subgraph of ${\Gamma_\beta(X,p)}$ where only the vertex $s_\sigma^{0b}$ is kept when $\sigma$ has a degenerate $1$-simplex in its boundary {together with t}he sign   defined by
\begin{equation}\label{eq:gamma p}
\gamma_p(x,\sigma) = \gamma(x,s_\sigma^{ab})
\end{equation}
where $s_\sigma^{ab}$ is the unique vertex with $p(s_\sigma^{ab})=0$ if $|\partial\sigma^\circ|=3$, or $s_\sigma^{0b}$ 
if $|\partial\sigma^\circ|=2$.
}
}

{The sign} $\gamma_p$ satisfies
\begin{equation}\label{eq:gamma-beta}
\prod_{x\in \partial\sigma^\circ} \gamma_p(x,\sigma) = 
\left\lbrace
\begin{array}{ll}
(-1)^{\beta(\sigma)} & |\partial\sigma^\circ|=3\\
(-1)^{a+\beta(\sigma)+1} &  |\partial\sigma^\circ|=2 ,\, x'=d_1\sigma\\
(-1)^{a+1} &  |\partial\sigma^\circ|=2,\, x'=d_0\sigma,\text{ or }d_2\sigma 
\end{array}
\right. 
\end{equation}
where $x'$ is the {unique} {degenerate} edge in the case $|\partial\sigma^\circ|=2$.

\Pro{\label{pro:reduction to signed graph}
Let $p\in \sDist_\beta(X)$ be a twisted distribution with no deterministic  simplices. 
Then
$$
\rank(p) = \rank(B( {\Gamma^0_\beta(X,p)} )).
$$ 
}

The choice we made in the definition of the graph does not affect the rank.

\begin{figure}[h!]
\centering
\begin{subfigure}{.49\textwidth}
  \centering
 \includegraphics[width=.4\linewidth]{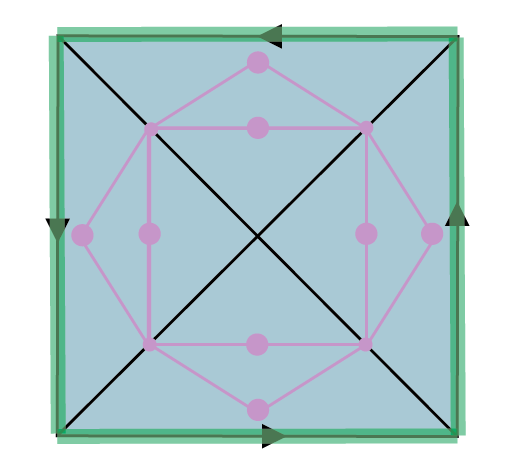}
  \caption{}
  \label{fig:Gamma}
\end{subfigure}%
\begin{subfigure}{.49\textwidth}
  \centering
 \includegraphics[width=.4\linewidth]{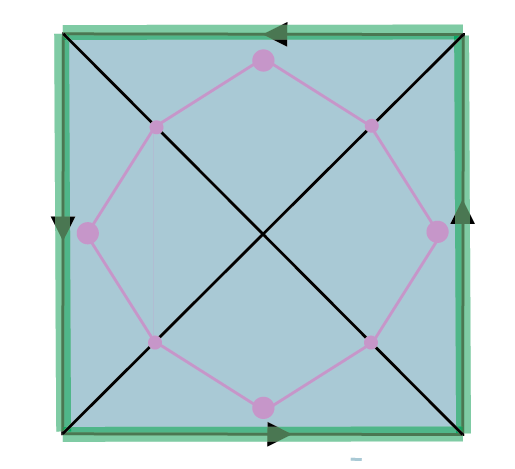}
  \caption{}
  \label{fig:Gamma-0}
\end{subfigure}
\caption{(a) The graph $\Gamma_{\bar\beta}(\bar X,\bar p)$ where $p$ is a PR box. (b) The subgraph graph $\Gamma_{\bar\beta}^0(\bar X,\bar p)$.}
\label{fig:CHSH-PR}
\end{figure}

\subsection{The rank formula}   
\label{sec:The rank formula}

We recall some basic definitions about signed graphs following  \cite{zaslavsky2013matrices}.  
Let $\Sigma=(V,E)$ be a simple (undirected) graph. 
A path is a sequence of edges $e_1e_2\cdots e_k$ with no repetition such that $e_{i-1}$ and $e_i$ has a common vertex. If $v$ and $w$ are the vertices of $e_1$ and $e_k$ not common to $e_2$ and $e_{k-1}$, respectively, then we say the path is from $v$ to $w$. A path is called closed if $v=w$. A circle is the graph determined by a closed path. If $\Sigma$
comes with a sign
$\gamma:E\to \set{\pm 1}$ then a circle $C$ given by the sequence $e_1e_2\cdots e_k$ has sign 
$$
\gamma(C) = \gamma(e_1)\gamma(e_2) \cdots \gamma(e_k).
$$
If $\gamma(C)=1$ ($\gamma(C)=-1$) the circle is {called} positive (negative).

\Def{\label{defbalanced}
{\rm
A signed graph $\Sigma$ is called balanced if every circle in it is positive. We will write $b(\Sigma)$ for the number of components of the graph that are balanced.
}}


A bidirected graph comes with a choice of sign $\eta(v,e)=\pm 1$ for each $v$ incident to an edge $e$. Given a signed graph $(\Sigma,\gamma)$ we can define a bidirected graph $(\Sigma,\eta)$ such that
$$
\gamma(e)=-\eta(v,e)\eta(w,e)
$$ 
for every edge $e=\set{v,w}$. This bidirected graph is not unique. An incidence matrix of $(\Sigma,\gamma)$ is a $|V|\times |E|$-matrix $H(\Sigma)$ such that  
$$
H(\Sigma)_{v,e} =\left\lbrace
\begin{array}{ll}
\eta(v,e) & v\in e\\
0 & \text{otherwise.}
\end{array}
\right.
$$

\Thm{\label{thm:rank incidence}
$H(\Sigma)$ has rank $|V|-b(\Sigma)$.
} 
\Proof{See \cite[Theorem IV.1]{zaslavsky2013matrices}.}

We will apply this rank result to  bipartite graphs.

\Def{\label{def:hat construction}
{\rm
Let $\Sigma$ be a signed bipartite graph with vertex set $V(\Sigma)=V^0(\Sigma)\sqcup V^1(\Sigma)$. Suppose that every vertex in $V^0(\Sigma)$ is incident to exactly two edges. Define a new bidirected graph $(\hat \Sigma,\eta)$ with
\begin{itemize}
\item vertex set $V^0(\Sigma)$,
\item edge set $V^1(\Sigma)$, and
\item bidirection defined by
$$
\eta(v,e) = \gamma(v,e)
$$
where $v\in V^0(\Sigma)$ and $e\in V^1(\Sigma)$ are connected by an edge in $\Sigma$.
\end{itemize}
}}


The vertices incident to an edge in $V^1(\Sigma)$ are those vertices in $V^0(\Sigma)$ connected to it by an edge in $\Sigma$. Effectively, to obtain $\hat \Sigma$ we merge any two edges incident to a vertex in {$V^0(\Sigma)$} into a single edge.
Then using Theorem \ref{thm:rank incidence} we have
\begin{equation}\label{eq:rank B}
\rank(B(\Sigma)) = \rank(H(\hat \Sigma))= |V^0(\Sigma)|-b(\hat \Sigma).
\end{equation}
Note that the rank does not depend on the choice of $\eta$ since a different choice would amount to multiplying the corresponding row with $-1$. 
If $\Gamma_\beta^0(X,p)$ satisfies the assumption of Definition \ref{def:hat construction} we can apply the construction $\Sigma\mapsto \hat \Sigma$ to the graph $\Gamma_\beta^0(X,p)$. We define
\begin{equation}\label{eq:balance-space}
b(X,p) = b( \hat \Gamma_\beta^0(X,p)  ).
\end{equation}

\Thm{\label{thm:main} 
Let $X$ be a simplicial set 
generated by   $2$-simplices $\sigma_1,\sigma_2,\cdots ,\sigma_N$ {such that
each $\partial\sigma_i$ consists of either three distinct non-degenerate {$1$-simplices} or two distinct non-degenerate {$1$-simplices} and a remaining degenerate {$1$-simplex}.}
{Consider}  
a twisted distribution $p\in \sDist_\beta(X)$ satisfying the following conditions:
\begin{itemize}
\item  for each generating $2$-simplex $\sigma$, $p_\sigma^{ab}=0$ for at least one pair $(a,b)\in \ZZ_2^2$, and
\item every non-degenerate $1$-simplex of $\bar X=X/Z_p$ belongs  precisely to two generating $2$-simplices.
\end{itemize}
Then we have
$$
\rank(p) = |(Z_p)_1^\circ| + |\bar X_2^\circ| - b(\bar X,\bar p).
$$ 
}

\Proof{ 
By Lemma \ref{lem:det-rank}, Proposition \ref{pro:reduction to signed graph}, and Equation (\ref{eq:rank B}) we have
$$
\rank(p) = |(Z_p)_1^\circ| + \rank(\Gamma_{\bar\beta}^0(\bar X,\bar p)) = |(Z_p)_1^\circ| + |X_2^\circ| - b(X,p).
$$
The assumption on $p$ implies that the number of the vertices of $\hat \Gamma_{\bar\beta}^0(\bar X,\bar p)$ is given by $|X_2^\circ|$.
} 

{Note that if the condition   $p_\sigma^{ab}=0$ fails for at least one pair of outcomes for every non-degenerate simplex  then to compute the rank we can restrict to the simplicial subset for which this condition holds.}

\section{Examples}
\label{sec:Examples}

The rank formula in Theorem \ref{thm:main} {is very useful in finding the vertices of the polytope of twisted distributions. 
We can partition $\sDist_\beta(X)$   by fixing a simplicial subset $Z\subset X$ and considering those twisted distributions $p$ with the property that  $Z_p=Z$.
Our approach will be to combine this observation with the following result.}
 
\Lem{\label{lem:Z_p 1 dimensional}
Let $p\in \sDist_\beta(X)$ be a vertex.
Then
$$
|(Z_p)_1^\circ|{-|(Z_p)_2^\circ|} \geq |X_1^\circ|-|X_2^\circ|.
$$
} 
\Proof{By Corollary \ref{cor:vertex} $p$ is a vertex if and only if $\rank(p)=|X_1^\circ|$. We have $\rank(\bar p)\leq |X_2^\circ|-{|(Z_p)_2^\circ|}$ since $Z_p$ is $1$-dimensional. Then the result follows from Lemma \ref{lem:det-rank}.
}

\begin{figure}[h!] 
  \centering
  \includegraphics[width=.2\linewidth]{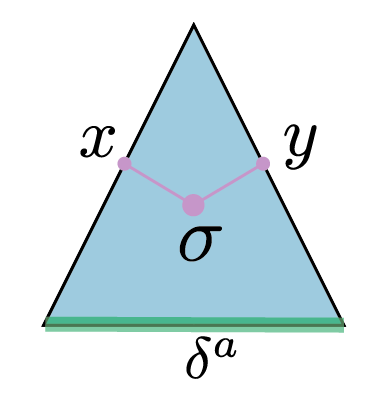}
\caption{
}
\label{fig:product of gamma}
\end{figure}

The rank formula contains the number of balanced components of the associated graph. 
For the computation of the signs of the circle graphs
we will use the following formula. 
See Figure (\ref{fig:product of gamma}).

\Lem{\label{lem:product of gamma}
Let $p\in \sDist_\beta(X)$ and consider $\gamma_p$ defined in Equation (\ref{eq:gamma p}). For a non-degenerate $2$-simplex $\sigma$ whose boundary $\partial\sigma=\set{x,y,z}$ contains a deterministic edge, say $z$, with $p_z=\delta^a$ we have
$$
\gamma_p(x,\sigma)\gamma_p(y,\sigma) = (-1)^{a+\beta(\sigma)+1}.
$$
}
\Proof{The distribution $p_\sigma\in D(\ZZ_2^2)$ is zero if and only if $(c,d)$ satisfies $c+d=a+\beta(\sigma)+1$. We have $\gamma(x,\sigma)=(-1)^c$ and $\gamma(y,\sigma)=(-1^d)$. Hence the formula follows. 
}

\subsection{$N$-cycle scenario}
\label{sec:N-cycle-scenario}

{For $N\geq 2$, l}et $\tilde C_N$ denote 
the measurement space of the $N$-cycle scenario defined as
the following simplicial set:
\begin{itemize}
\item Generating $2$-simplices: $\sigma_1,\cdots,\sigma_N$.
\item Identifying relations:
$$
d_{i'_1}\sigma_1 = d_{i_2}\sigma_2,\; d_{i_2'}\sigma_2= d_{i_3}\sigma_3\;\cdots\; d_{i'_N} \sigma_N = d_{i_1} \sigma_1 
$$
where $i_k\neq i_k' \in {\set{0,1}}$ for $1\leq k\leq N$.
\end{itemize}  
{The case $N=4$ where the boundary is oriented in the counter-clockwise direction is depicted in Figure (\ref{fig:CHSH}).}
{See also \cite[Definition 4.1]{kharoof2023topological}.} 
The edges on the boundary will be denoted by $x_1,x_2,\cdots,x_N$.

 \begin{figure}[h!]
\centering
\begin{subfigure}{.49\textwidth}
  \centering
  \includegraphics[width=.4\linewidth]{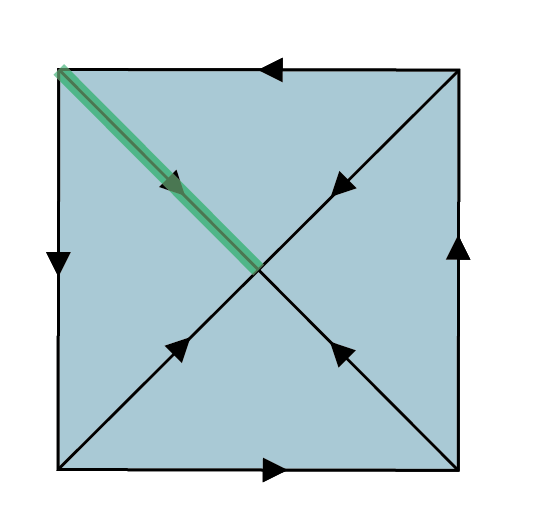}
  \caption{}
  \label{fig:interior}
\end{subfigure}%
\begin{subfigure}{.49\textwidth}
  \centering
  \includegraphics[width=.4\linewidth]{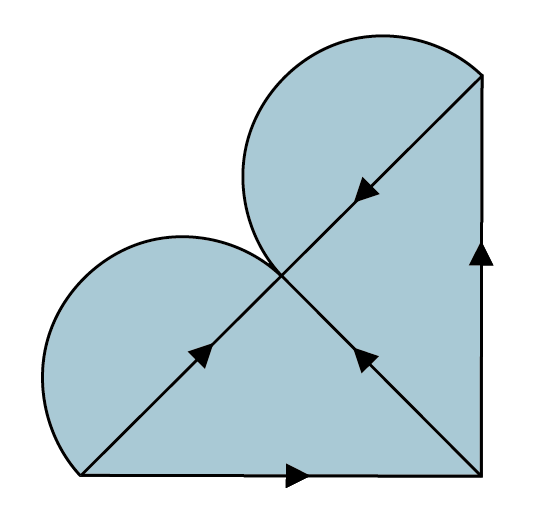}
  \caption{}
  \label{fig:line}
\end{subfigure}
\caption{(a) Deterministic interior edge. (b) Collapsed scenario.
}
\label{fig:interior-collapsed}
\end{figure}

Let $p\in \sDist_\beta(\tilde C_N)$ be  a vertex. 
Let us consider the possibilities for $Z_p$.
\begin{enumerate}
\item $Z_p$ has a deterministic edge connecting a boundary vertex to the central vertex: In this case the quotient $\tilde C_N/Z_p$ can be identified with a wedge sum of $D_{K_i}$'s of Example \ref{ex:line} where $K_i\leq N$.
We know that every twisted distribution on $D_{K_i}$ is non-contextual.
Then Corollary \ref{cor:quotient non-contextual} implies that $p$ is non-contextual, {hence a deterministic distribution}. The situation is depicted in Figure (\ref{fig:interior-collapsed}).


\item All deterministic edges of $Z_p$ are on the boundary:  
Note that by the previous case we know that $p$ would be a deterministic distribution if $Z_p$ contained a deterministic triangle. 
Assuming that 
$Z_p$ does not contain a deterministic triangle,  Lemma \ref{lem:Z_p 1 dimensional} gives 
$$
|(Z_p)_1^\circ| \geq |X_1^\circ| - |X_2^\circ| = 2N-N=N.
$$
This forces $Z_p$ to be the boundary $\partial \tilde C_N$ of the disk. 
Assume that $p|_{\partial \tilde C_N}=\delta^s$. Then  we have
$$
\delta^s_{x_i} = \delta^{a_i},\;\;\; i=1,2,\cdots,N
$$
for some $a_i\in \ZZ_2$. 
The graph $\hat \Gamma_{\bar\beta}^0(\bar X,\bar p)$ 
is given by a circle $C$ with edges $e_1,e_2,\cdots,e_N$ as in Figure (\ref{fig:Gamma-hat}). The sign of the circle is given by 
$$
\gamma_{\bar p}(C) = \prod_{i=1}^N \gamma_{\bar p}(e_i) = (-1)^{N} \prod_{i=1}^N (-1)^{a_i+\beta(\sigma_i)+1} =(-1)^N\prod_{i=1}^N (-1)^{\bar\beta(\sigma_i)+1} =  (-1)^{[\bar\beta]}
$$
and by the rank formula we obtain
\begin{equation}\label{eq:rank CN}
\rank(p) = \left\lbrace
\begin{array}{ll}
2N & [\bar \beta]=1 \\
2N-1 & [\bar \beta]=0.
\end{array}
\right.
\end{equation}
Therefore $p$ is a vertex if and only if $[\bar\beta]=1$, that is, $[\beta]=1+\sum_{i=1}^N a_i$.
\end{enumerate} 


As a special case let us take $[\beta]=0$.
In this case $p$ is a vertex if and only if $\sum_{i=1}^N a_i=1$. 
When $N=4$ this reproduces the PR boxes in Example \ref{ex:CHSH}. 
 

 \begin{figure}[h!]
\centering
\begin{subfigure}{.49\textwidth}
  \centering
  \includegraphics[width=.4\linewidth]{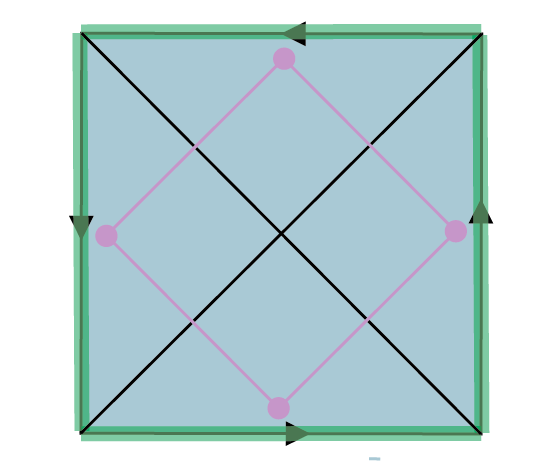}
  \caption{}
  \label{fig:Gamma-hat}
\end{subfigure}%
\begin{subfigure}{.49\textwidth}
  \centering
  \includegraphics[width=.5\linewidth]{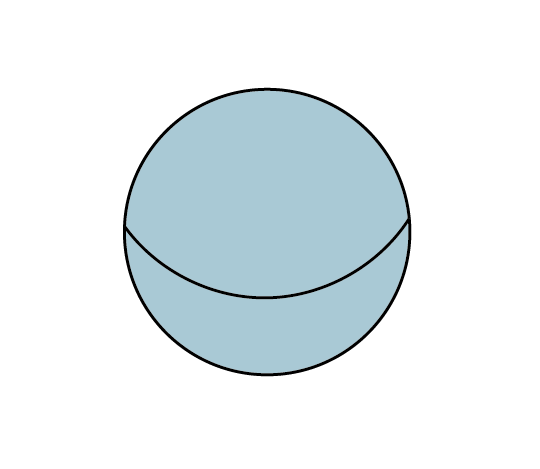}
  \caption{}
  \label{fig:sphere}
\end{subfigure}
\caption{(a) The graph $\hat \Gamma_{\bar \beta}^0( \tilde C_4/\partial\tilde C_4)$. (b) When the boundary is collapsed the resulting space is a sphere.
}
\label{fig:Gamma-hat-sphere}
\end{figure}

\subsection{{Boundary of tetrahedron}}
\label{sec:boundary of tetrahedron}

Let $\partial\Delta^3$ denote the simplicial set given by the boundary of a tetrahedron with generating $2$-simplices $\sigma_1,\sigma_2,\sigma_3,\sigma_4$. Let $p\in \sDist_\beta(\partial\Delta^3)$ {be a vertex}. 
Then Lemma \ref{lem:Z_p 1 dimensional} gives
$$
|(Z_p)_1^\circ|- |(Z_p)_2^\circ| \geq 4-2=2
$$
and we have
$|(Z_p)_2^\circ|=0,1$ or $4$. 
The last case gives a deterministic distribution if $[\beta]=0$. Let us consider the remaining cases.
\begin{enumerate}
\item  $|(Z_p)_2^\circ|=0$:  In this case ${\hat\Gamma_{\bar\beta}^0(\bar X,\bar p)}$ is a circle $C$ with edges $e_1,e_2,e_3,e_4$; see Figure (\ref{fig:t1}). Its sign is given by
$$
\gamma_{\bar p}(C) = \prod_{i=1}^4\gamma_{\bar p}(e_i) = (-1)^{a+\beta(\sigma_1)+1}(-1)^{b+\beta(\sigma_2)+1}(-1)^{a+\beta(\sigma_3)+1}(-1)^{b+\beta(\sigma_4)+1} = (-1)^{[\beta]}
$$
and by the rank formula
\begin{equation}\label{eq:rank tetrahedron}
\rank(p) = \left\lbrace
\begin{array}{ll}
6 & [ \beta]=1 \\
5 & [\beta]=0.
\end{array}
\right.
\end{equation}
Therefore $p$ is a vertex if and only if $[\beta]=1$.

\item $|(Z_p)_2^\circ|=1$:  This case (Figure (\ref{fig:t2})) reduces to the cycle scenario $\tilde C_3$ where $p|_{\partial \tilde C_3}$ is deterministic with $a+b+c=0$. Therefore by Equation (\ref{eq:rank CN}), $p$ is a vertex if and only if $[\beta]=1$.
\end{enumerate}
In the case $[\beta]=0$ we observe that all the vertices are deterministic, which {provides} an alternative approach to the computation in \cite[Proposition 4.12]{okay2022simplicial}.


 \begin{figure}[h!]
\centering
\begin{subfigure}{.49\textwidth}
  \centering
  \includegraphics[width=.4\linewidth]{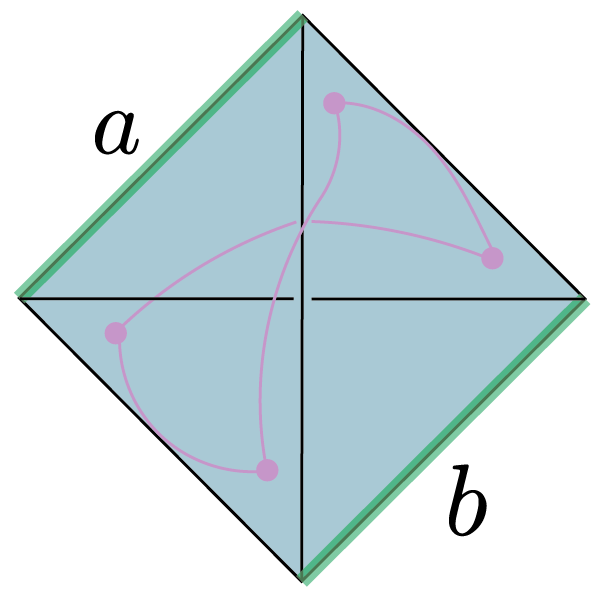}
  \caption{}
  \label{fig:t1}
\end{subfigure}%
\begin{subfigure}{.49\textwidth}
  \centering
  \includegraphics[width=.4\linewidth]{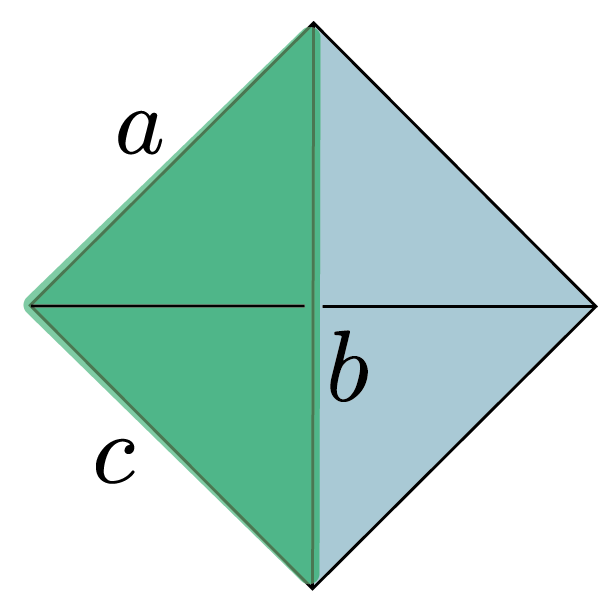}
  \caption{}
  \label{fig:t2}
\end{subfigure}
\caption{
}
\label{fig:boundary-tetrahedron}
\end{figure}

\subsection{Mermin torus}
\label{sec:mermin torus}

Let $T$ denote the Mermin torus introduced in \cite{okay2022mermin}. 
Let $p\in \sDist_\beta(T)$ be a vertex. The distribution $p$ is deterministic if and only if $[\beta]=0$.
Let us assume $p$ is not deterministic. Lemma \ref{lem:Z_p 1 dimensional} gives
$$
|(Z_p)_1^\circ| - |(Z_p)_2^\circ|\geq 3
$$ 
which can only be satisfies if $Z_p$ consists of three generating $1$-simplices as in Figure (\ref{fig:Mermin-a}) or two generating triangles as in Figure (\ref{fig:Mermin-b}). In both cases 
$\hat\Gamma_{\bar\beta}^0(\bar X,\bar p)$
is a circle $C$. We compute the sign of this circle for each case:
\begin{enumerate}
\item In Figure (\ref{fig:Mermin-a}) we have
$$ 
\gamma_{\bar p}(C) = (-1)^{a+\beta(\sigma_1)+1}(-1)^{b+\beta(\sigma_2)+1}(-1)^{c+\beta(\sigma_3)+1}
(-1)^{a+\beta(\sigma_4)+1}(-1)^{b+\beta(\sigma_5)+1}(-1)^{c+\beta(\sigma_6)+1} =(-1)^{[\beta]}.
$$
Therefore $p$ is a vertex if and only if $[\beta]=1$.

\item In Figure (\ref{fig:Mermin-b}) we have
$$ 
\gamma_{\bar p}(C) = (-1)^{a+\beta(\sigma_1)+1}(-1)^{b+\beta(\sigma_3)+1}(-1)^{a+b+c+\beta(\sigma_2)+\beta(\sigma_5)+\beta(\sigma_4)+1}
(-1)^{c+\beta(\sigma_6)+1} =(-1)^{[\beta]}
$$
and $p$ is a vertex if and only if $[\beta]=1$.
\end{enumerate}
Combining these two observations gives the main result of \cite{okay2022mermin}. $\sDist_\beta(T)$ has only deterministic vertices if $[\beta]=0$. On the other hand, if $[\beta]=1$ then 
there are two kinds of vertices as given in Figure (\ref{fig:Mermin}).

 \begin{figure}[h!]
\centering
\begin{subfigure}{.49\textwidth}
  \centering
  \includegraphics[width=.5\linewidth]{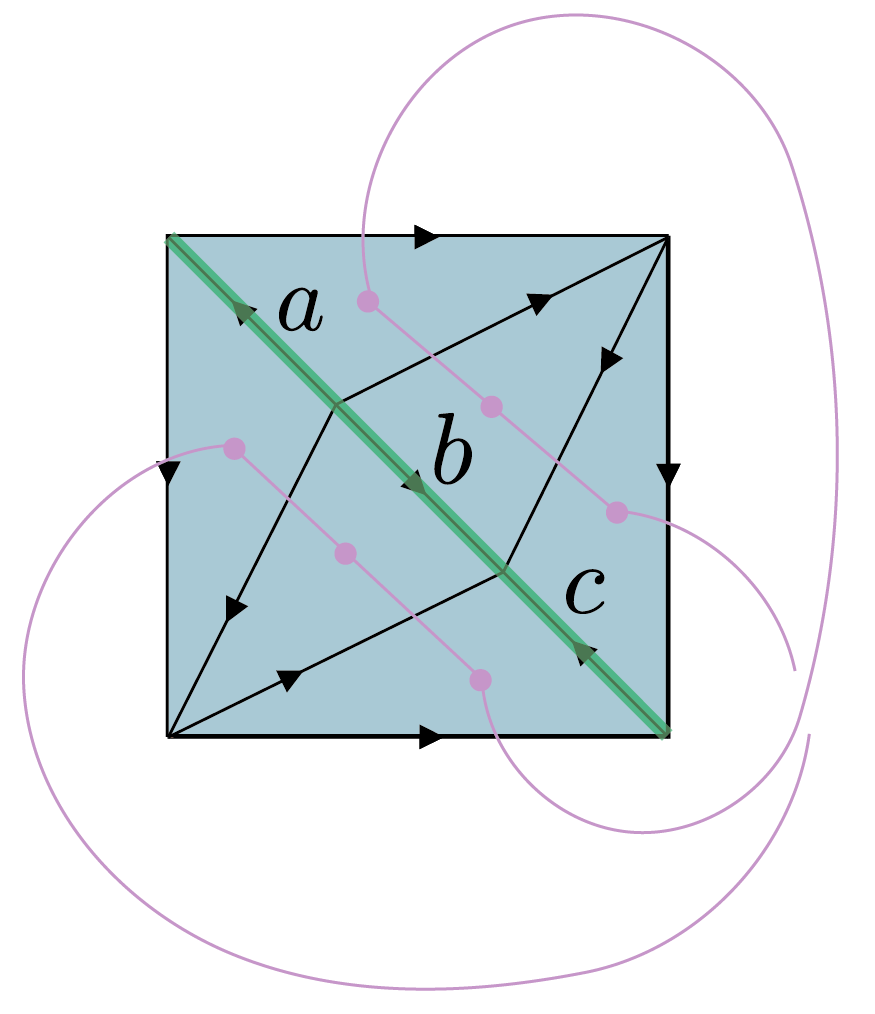}
  \caption{}
  \label{fig:Mermin-a}
\end{subfigure}%
\begin{subfigure}{.49\textwidth}
  \centering
  \includegraphics[width=.5\linewidth]{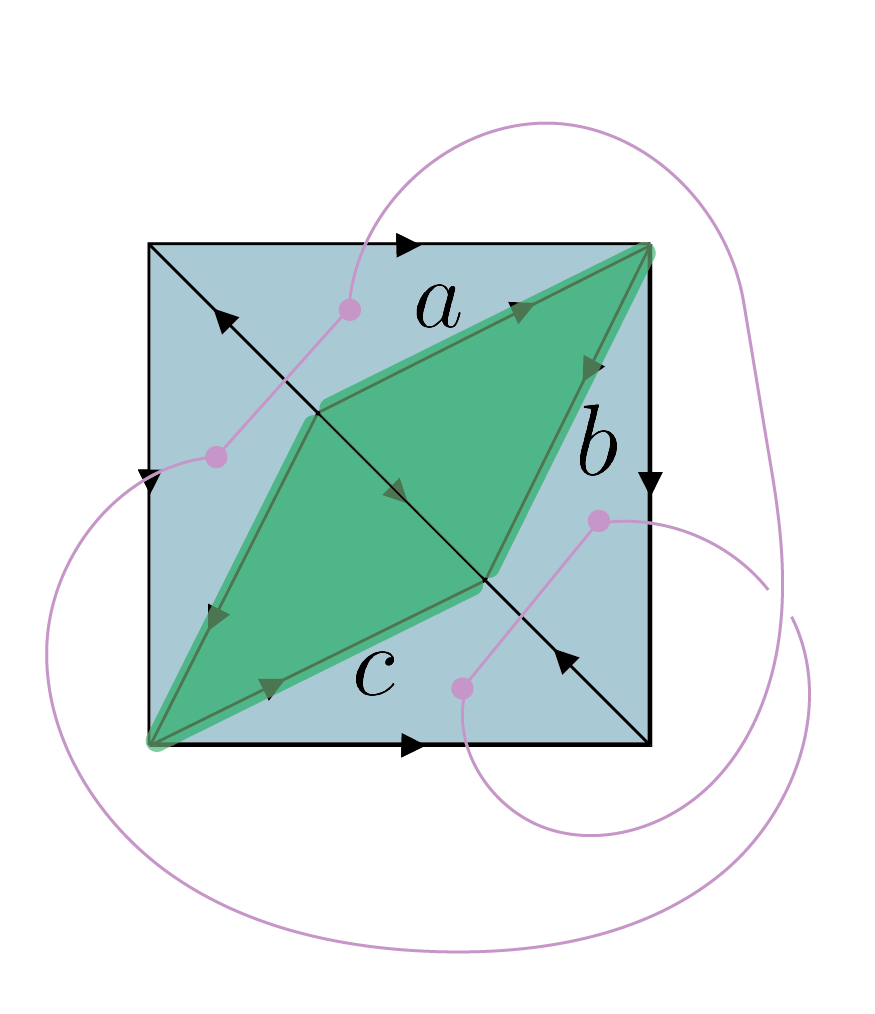}
  \caption{}
  \label{fig:Mermin-b}
\end{subfigure}
\caption{
}
\label{fig:Mermin}
\end{figure}

\bibliography{bib.bib}
\bibliographystyle{ieeetr}

\end{document}